\documentclass[prd,nofootinbib,a4paper]{revtex4}

\usepackage{graphicx,epsfig,amssymb,subfigure}
\usepackage[latin1]{inputenc}

\def\be {\begin{equation}}
\def\ee {\end{equation}}
\def\ba {\begin{eqnarray}}
\def\ea {\end{eqnarray}}
\def\nn {\nonumber}
\def\bea{\begin{eqnarray}}
\def\eea{\end{eqnarray}}
\def\no {\noindent}
\def\bi {\begin{itemize}}
\def\ei {\end{itemize}}

\begin{document}
\vspace{5mm}

\title{Area spectrum of slowly rotating black holes}

\vspace{5mm}

\author{Yun Soo Myung}
\email{ysmyung@inje.ac.kr}
\affiliation{ Institute of Basic Science
and School of Computer Aided Science \\ Inje University, Gimhae
621-749, Korea}

\vspace{5mm}

\begin{abstract}
We investigate   the area spectrum for  rotating black holes which
are Kerr and BTZ black holes. For slowly rotating black holes, we
use the Maggiore's idea combined with  Kunstatter's method to derive
their area spectra, which are equally spaced.
\\

\end{abstract}
 \maketitle

\par\noindent

\section{Introduction}
It is believed that black holes could  provide a test bed for any
proposed scheme for a quantum theory of gravity. Hod  has  combined
the perturbations of  black holes with the  quantum mechanics and
statistical physics in order to derive the quantum of the black hole
area spectrum~\cite{Hod}. For highly excited Schwarzschild black
hole, Hod has used the real part of quasinormal frequencies (QNFs)
to derive the area quanta $\Delta A= 4\ln(3)l_p^2$. However, it is
not consistent with  $\Delta A= 8\pi l_p^2$  which was obtained by
Bekenstein~\cite{Bek} from  the fact that the black hole area is
adiabatically invariant. Kunstatter has shown that the area spectrum
is equally spaced for $d\ge4$ dimensional Schwarzschild black holes
by using the adiabatically invariant integral~\cite{Kun}
 \be \label{adiabatic} I=\int \frac{dE}{\omega(E)} \to \int
\frac{dM}{\omega_R},\ee where $(E,\omega)$ are  (energy, vibrational
frequency) and $(M,\omega_R)$ are (black hole mass, real part of
QNFs). On later, Maggiore has proposed that a black hole perturbed
by external field is considered as a collection of damped harmonic
oscillators~\cite{Magg}. Accordingly, he regarded
$\omega_0=\sqrt{\omega_R^2+\omega_I^2}$ as a physically proper
frequency and thus, $\omega_0= \omega_I$ was used to derive $\Delta
A= 8\pi l_p^2$  for highly excited QNFs of $\omega_I \gg \omega_R$
by considering the transition $n\to n-1$.

For Kerr black holes, their area spectra  were not clearly
determined because of their spinning: these solutions are stationary
but not static~\cite{SV,Vag}. However, it was argued that for $J \ll
M^2$, area spectrum of Kerr black hole could be obtained and thus,
is equally spaced~\cite{Med}.  Kerr black holes are the most
interesting black holes in astronomical point of view, while in
theoretical point of view they are more complicated than
Schwarzschild  and Reissner-Nordstr\"om black holes.  Also, the BTZ
black hole is important because it is closely related to the Kerr
black hole even it belongs to a toy model for studying rotating
black holes~\cite{BTZ}.

In this work, we will show that area spectrum could be consistently
derived only for slowly rotating black
holes~\cite{HH,Po,Mar,GS,KC,Fer}. Generally, ``slowly rotating (sr)"
means that one may consider up to linear order of rotating parameter
$a=J/M(a \ll 1)$ in the metric functions, equations of motion, and
thermodynamic quantities. Further, the slowly rotating black hole
must be far away from extremality because the extremal black hole
means $J=M^2(a/M=1)$. Thus, surface gravity (temperature) and area
of event horizon do not change to ${\cal O} (a)$. However, we
consider the angular momentum $J$ at this order and thus, angular
velocity of $\Omega_+ \sim J$ at the event horizon.

\section{Slowly rotating Kerr black hole}
The metric of a four-dimensional Kerr black hole given in
Boyer-Lindquist coordinates is \bea
ds^{2}_{Kerr}&=&-(1-\frac{2Mr}{\Sigma})dt^{2}-\frac{4Mar
\sin^{2}\theta}{\Sigma}dtd\varphi+\frac{\Sigma}{\tilde{\Delta}}dr^{2}\nn\\
&+&\Sigma d\theta^{2}+(r^2+a^2+2Ma^2r\sin^{2}\theta) \sin^{2}\theta
d\varphi^{2} \label{met} \eea where, as always, $M$ is the mass of
the black hole, $J$ is the angular momentum of the black hole,
$a=J/M$ is the specific angular momentum, $\Sigma=r^2+a^2\cos^2
\theta$, and $\tilde{\Delta}=r^2-2Mr+a^2$. The roots of
$\tilde{\Delta}=0$ from $g^{rr}=0$ are given by \be r_{\pm}=M\pm
\sqrt{M^{2}-a^{2}} \label{root} \ee where $r_{+}$ is the radius of
the event (outer) black hole horizon and $r_{-}$ is the radius of
the inner black hole horizon. In the case of slowly rotating case,
we have approximate horizon radii\be r_+ \approx 2M
-\frac{J^2}{2M^3},~~r_-\approx \frac{J^2}{2M^3}\ee which satisfy a
relation of  $r_++r_-\approx 2M$.
 The Kerr
black hole is rotating with angular velocity evaluated at the event
horizon \be
\Omega_+=\frac{J}{2M\left(M^{2}+\sqrt{M^{4}-J^{2}}\right)} \approx
\frac{J}{4M^3}\equiv \Omega_+^{sr}, \label{angve} \ee where the last
expression ($\approx$) means angular velocity for slowly rotating
Kerr black hole.  Furthermore, the horizon area and the Hawking
temperature of Kerr black hole (in gravitational units of $c=G=1$
and $\hbar=l^2_p$) are given, respectively, by \bea
A&=&8\pi\left(M^{2}+\sqrt{M^{4}-J^{2}}\right)\approx
8\pi\Big(2M^2-\frac{a^2}{2}\Big)\equiv A^{sr} \label{area} \eea and
\bea T_{H}&=&\frac{\hbar \sqrt{M^{4}-J^{2}}}{4\pi
M\left(M^{2}+\sqrt{M^{4}-J^{2}}\right)}\approx \frac{\hbar}{8\pi
M}\equiv T^{sr}_H. \label{temeq} \eea For slowly rotating black
hole,  we keep up  $a^2=J^2/M^2$-term in the horizon area but drop
off $a^2/M^2$-term in the Hawking temperature because the latter is
very smaller than the former.  If one keeps the first term in
$A^{sr}$ only, one could not get slowly rotating black hole.

Before we proceed, we mention QNFs given by~\cite{KHod,KNei} \be
\omega_n=\tilde{\omega}_0-i\Big[4\pi T_0(a)(n+1/2)\Big]\ee with the
effective temperature $T_0(a)$. For slowly rotating black hole, one
finds that $T_0(a=0)\approx -T^{sr}_H/2$. Then, we note that the
transition frequency is determined by~\cite{Vag} \be
\hbar\omega_c=\hbar\Big[(\omega_I)_n-(\omega_I)_{n-1}\Big]=2\pi
T^{sr}_H=\frac{\hbar}{4M}. \ee

We are in a position to derive the area spectrum of  slowly rotating
Kerr black hole by employing Kunstatter's method~\cite{Kun}.
Implementing the first law of slowly rotating black hole \be
 dM=T^{sr}_HdS^{sr}+\Omega_+^{sr}dJ, \ee the
adiabatically invariant integral (\ref{adiabatic}) is given as \bea
I^{sr}_{Kerr}&=&\int \frac{dM -\Omega_+^{sr} dJ}{\omega_c}\label{adiabkerr1}\\
&=&\int\Big[4MdM-\frac{J~dJ}{M^2}\Big] \eea which is consistent with
the leading-order terms in~\cite{Med}.
 The adiabatically  invariant integral  leads to \be
I^{sr}_{Kerr}= 2M^2-\frac{1}{2}\frac{J^2}{M^2}=\frac{A^{sr}}{8\pi},
\label{adiabkerr2} \ee which is obviously correct when considering
the adiabatically invariant integral as
 \be
I^{sr}_{Kerr}=\int \frac{T_H^{sr} dS^{sr} }{\omega_c}
=\frac{1}{8\pi} \int dA^{sr}\ee with the entropy
$S^{sr}=\frac{A^{sr}}{4\hbar}$. Using the Bohr-Sommerfeld
quantization condition of $I^{sr}_{Kerr} \approx n \hbar$, the
quantized area spectrum is \be \mathcal{A}^{sr}_n=8\pi \hbar n
\label{area2}
 \ee
which is the universal area spectrum. On the other hand, one has
found $\mathcal{A}^{sr}_n=4 \hbar n$ when using the tunneling
method~\cite{BMV}. The entropy spectrum takes the form \be
S^{sr}_n=\frac{\mathcal{A}^{sr}_n}{4\hbar}=2\pi n. \ee In the case
of $J\to 0$, we immediately obtain the area spectrum for the
Schwarzschild black hole. On the other hand, for Kerr black hole,
the adiabatically invariant integral is given by~\cite{Vag,Med} \be
I_{Kerr}=\int\Big[4MdM-\frac{2JdJ}{M^2+\sqrt{M^4-J^2}}\Big]
=\frac{A}{4\pi}-2M^2 \ln\Big[\frac{A}{8\pi}\Big] \ee where the
logarithmic term becomes dominant, which gives rise to difficulty to
interpret the entropy.

\section{slowly rotating BTZ black hole}

In this section, we study the slowly rotating BTZ black hole in
gravitational units of $c=8G=1$. In three dimensional spacetimes,
the metric of the BTZ black hole is given by~\cite{BTZ} \be
ds^2_{BTZ} =-\left(-M+{ r^2 \over l^2} +{ J^2 \over {4\,r^2} }
\right)dt^2+ { {dr^2} \over {\left(-M+{ r^2 \over l^2} +{ J^2 \over
{4\,r^2} }\right)} } +\, r^2 \, \left( d\phi -{J \over {2\,r^2}
}\,dt\right) ^2 ~, \ee where the cosmological constant is given by
$\Lambda=-{1 \over l^2}$. From the condition of $g^{rr}=0$, we
obtain the outer and inner horizons at \be \label{hori1}
r^2_{\pm}=\frac{l^2M}{2}\Bigg[1\pm
\sqrt{1-\frac{J^2}{l^2M^2}}\Bigg]. \ee The mass $M$ and angular
momentum $J$ of the black hole can be expressed in terms of  $r_\pm$
as
\begin{equation}
M= { {  {r_+ ^2}+ {r_- ^2} } \over l^2},~~ J= { {  2\,{r_+}\, {r_-}
} \over l} ~.
\end{equation}
Using these expressions, we rewrite two horizons as \be
\label{hori2} r_\pm=\frac{l}{2}\Bigg[\sqrt{M+\frac{J}{l}}\pm
\sqrt{M-\frac{J}{l}}\Bigg]. \ee  The slowly rotating BTZ black hole
implies that \be a=\frac{J}{M} \ll 1. \ee Then, from (\ref{hori1})
and (\ref{hori2}),  we have approximate two horizons  \be r_+
\approx l\sqrt{M}\Big[1-\frac{1}{8} \Big(\frac{J}{l M}\Big)^2
\Big],~~r_-\approx \frac{J}{2\sqrt{M}}. \ee
 The BTZ
black hole is rotating with angular velocity evaluated at the event
 horizon \be
\Omega_+=\frac{J}{2r_+^2}=\frac{J}{l^2\left(M^{2}+\sqrt{M^{2}-\frac{J^{2}}{l^2}}\right)}
\approx \frac{J}{2l^2M}\equiv \Omega_+^{sr}, \label{angveB} \ee
where the last expression ($\approx$) means angular velocity for
slowly rotating BTZ black hole. Also, the horizon area and the
Hawking temperature of BTZ black hole  are given, respectively,
by~\cite{BTZ,CC} \bea A&=&2\pi r_+\approx 2\pi
l\sqrt{M}\Big[1-\frac{1}{8} \Big(\frac{J}{l M}\Big)^2 \Big] \equiv
A^{sr}\label{areaB} \eea and \bea T_{H}&=&\frac{\hbar
(r_+^2-r_-^2)}{2\pi l^2r_+}\approx \frac{\hbar \sqrt{M}}{2\pi l
}\equiv T^{sr}_H. \label{temeq} \eea For slowly rotating BTZ black
hole with $l\gg 1$, we keep up $\frac{a^2}{l}=\frac{J^2}{lM^2}$-term
in the horizon area but drop off
$\frac{a^2}{l^3}=\frac{J^2}{l^3M^2}$-term in the Hawking
temperature.

The two types of  quasinormal modes of the  BTZ black hole for a
massive scalar field are given by~\cite{CL,Bir,LKM}
\begin{eqnarray}
\omega _{R} &=& -{\frac {m}{l}}- i {\frac { (r_{+}+r_{-} ) \left( 2
n+1+{\sqrt {1+\mu \,} }
 \right) }{{l}^{2}}} ~,\\
\omega _{L} &=& {\frac {m}{l}}- i {\frac { (r_{+}-r_{-} ) \left( 2
n+1+{\sqrt {1+\mu \,} }
 \right) }{{l}^{2}}}   ~,
\end{eqnarray}
where $m$ and $ n$ are the angular quantum number and the overtone
quantum number respectively.  $\mu$ is the mass parameter defined by
$\mu \equiv m^2\,l^2 / {\hbar ^2}$, where $m$ is the mass of the
scalar field.  At large $n$ for a fixed $\vert m \vert$ ($n \gg
\vert m \vert$), the two  transition frequencies of $\omega _{Rc}$
and $ \omega _{Lc}$ are given by~\cite{KN}
\begin{eqnarray}
\omega _{R c} &=&  {{2\, (r_{+}+r_{-} ) } \over {l^2}} = { {2 \sqrt{M+{J / l} \, } } \over l} ~~,\\
\omega _{L c} &=&  {{2 \,(r_{+}-r_{-} ) } \over {l^2}} = { {2
\sqrt{M-{J / l} \, } } \over l}  ~.
\end{eqnarray}
For the slowly rotating BTZ black hole, however, we have one-type as
\be \label{app-omega}\omega _{R c} \approx  \omega _{L c} \approx
\frac{2\sqrt{M}}{l} =\frac{4\pi T^{sr}_H}{\hbar}.\ee The
adiabatically invariant integral is calculated to be  \bea
I^{sr}_{BTZ}&=&\int \frac{dM-\Omega^{sr}dJ}{\omega_{Rc}} \\ \nn
&=&l\sqrt{M}\Big[1-\frac{1}{8}\Big(\frac{J}{l M}\Big)^2\Big] \\
\nn &=&r_+ \\ \nn &=& \frac{A^{sr}}{2\pi}. \eea For 3D slowly
sinning  dilaton black hole, Fernando has obtained a similar
result~\cite{Fer}, but he has  neglected $J^2$-term to derive area
spectrum. Also, Kwon and Nam have used $J_R=\int dM/\omega_{Rc}$ and
$J_L=\int dM/\omega_{Lc}$ to derive two area spectra~\cite{KN}. In
this case, however, it is  questionable that  $J_R$ and $J_L$ could
represent the area (entropy) of the BTZ black hole  because they did
not take into account $\Omega dJ$ seriously. Without
$\Omega^{sr}dJ$, the first law of thermodynamics is not satisfied
for slowly rotating BTZ black hole.

 Applying  the
Bohr-Sommerfeld  condition of $I^{sr}_{BTZ} \approx n \hbar$ to the
slowly rotating BTZ black hole, the quantized area spectrum is
determined to be  \be \mathcal{A}^{sr}_n=2\pi \hbar n \label{areaB}
 \ee
which is the universal area spectrum. Considering  the unit of
$G=1/8$, the entropy spectrum takes the form \be
S^{sr}_n=\frac{2\mathcal{A}^{sr}_n}{\hbar}=4\pi n. \ee In the case
of $J\to 0$, we obtain the area spectrum for the non-rotating  black
hole~\cite{Set,LM}. On the other hand, for BTZ black hole, the
adiabatically invariant integral take a complicated form  when using
(\ref{app-omega}) \bea
I_{BTZ}&=&l\int\Bigg[\frac{dM}{2\sqrt{M}}-\frac{JdJ}{l^2\sqrt{M}\Big(M+
\sqrt{M^2-\frac{J^2}{l^2}}\Big)}\Bigg] \\ \no
&=&l\sqrt{M}+\frac{l}{2\sqrt{M}}
\sqrt{M^2-\frac{J^2}{l^2}}-\frac{l\sqrt{M}}{2}
 \ln
\Big[M+\sqrt{M^2-\frac{J^2}{l^2}}\Big],  \eea where the logarithmic
term becomes dominant. It implies that the presence of logarithmic
term  gives rise to difficulty in order to interpret $I_{BTZ}$ as
the entropy.
\section{Discussion}
We have investigated   the area spectrum for  rotating black holes
such as Kerr and BTZ black holes.  Rotating black holes provided the
ill-defined (adiabatically invariant) integral which contains a
large logarithmic term without any quantum corrections. This means
that Kunstatter's method to does not work for rotating black holes
well. However, for slowly rotating black holes, we have used the
Maggiore's idea combined with Kunstatter's method to derive their
area spectra, which are equally spaced. This shows clearly  that
slowly rotating black holes provide the quantum of area spectrum
very well, like the static black hole of Schwarzschild black hole.

{\it Acknowledgments}--  This work was in part  supported  by Basic
Science Research Program through the National Research  Foundation
(NRF) of Korea funded by the Ministry of Education, Science and
Technology (2009-0086861) and   the NRF grant funded by the Korea
government(MEST) through the Center for Quantum Spacetime (CQUeST)
of Sogang University with grant number 2005-0049409.



\begin{thebibliography}{X}




  \bibitem{Hod}
  S.~Hod,
  Phys.\ Rev.\ Lett.\  {\bf 81}, 4293 (1998)
  [arXiv:gr-qc/9812002].



\bibitem{Bek}
  J.~D.~Bekenstein,
  Phys.\ Rev.\  D {\bf 7}, 2333 (1973).

\bibitem{Kun}
 G.~Kunstatter,
  Phys.\ Rev.\ Lett.\  {\bf 90}, 161301 (2003)
  [arXiv:gr-qc/0212014].

\bibitem{Magg}
  M.~Maggiore,
  Phys.\ Rev.\ Lett.\  {\bf 100}, 141301 (2008)
  [arXiv:0711.3145 [gr-qc]].

\bibitem{SV}
  M.~R.~Setare and E.~C.~Vagenas,
  Mod.\ Phys.\ Lett.\  A {\bf 20}, 1923 (2005)
  [arXiv:hep-th/0401187].

\bibitem{Vag}
  E.~C.~Vagenas,
  JHEP {\bf 0811}, 073 (2008)
  [arXiv:0804.3264 [gr-qc]].


\bibitem{Med}
  A.~J.~M.~Medved,
  Class.\ Quant.\ Grav.\  {\bf 25}, 205014 (2008)
  [arXiv:0804.4346 [gr-qc]].


\bibitem{BTZ}
M.~Banados, C.~Teitelboim and J.~Zanelli,
  Phys.\ Rev.\ Lett.\  {\bf 69}, 1849 (1992)
  [arXiv:hep-th/9204099].

\bibitem{HH}
  J.~H.~Horne and G.~T.~Horowitz,
  Phys.\ Rev.\  D {\bf 46}, 1340 (1992)
  [arXiv:hep-th/9203083].

  \bibitem{Po}
  E.~Poisson,
  Phys.\ Rev.\  D {\bf 48}, 1860 (1993).

\bibitem{Mar}
  E.~A.~Martinez,
  Phys.\ Rev.\  D {\bf 50}, 4920 (1994)
  [arXiv:gr-qc/9405033].


  \bibitem{GS}
  T.~Ghosh and S.~SenGupta,
  Phys.\ Rev.\  D {\bf 76}, 087504 (2007)
  [arXiv:0709.2754 [hep-th]].


\bibitem{KC}
  H.~C.~Kim and R.~G.~Cai,
  Phys.\ Rev.\  D {\bf 77}, 024045 (2008)
  [arXiv:0711.0885 [hep-th]].

\bibitem{Fer}
  S.~Fernando,
  Phys.\ Rev.\  D {\bf 79}, 124026 (2009)
  [arXiv:0903.0088 [hep-th]].




\bibitem{KHod}
  U.~Keshet and S.~Hod,
  Phys.\ Rev.\  D {\bf 76}, 061501 (2007)
  [arXiv:0705.1179 [gr-qc]].


\bibitem{KNei}
  U.~Keshet and A.~Neitzke,
  Phys.\ Rev.\  D {\bf 78}, 044006 (2008)
  [arXiv:0709.1532 [hep-th]].

  \bibitem{BMV}
  R.~Banerjee, B.~R.~Majhi and E.~C.~Vagenas,
  Phys.\ Lett.\  B {\bf 686}, 279 (2010)
  [arXiv:0907.4271 [hep-th]].



  \bibitem{CC}
  R.~G.~Cai and J.~H.~Cho,
  Phys.\ Rev.\  D {\bf 60}, 067502 (1999)
  [arXiv:hep-th/9803261].

\bibitem{ML}
  Y.~S.~Myung and H.~W.~Lee,
  Mod.\ Phys.\ Lett.\  A {\bf 21}, 1737 (2006)
  [arXiv:hep-th/0506031].



\bibitem{CL}
V.~Cardoso and J.~P.~S.~Lemos,
  Phys.\ Rev.\  D {\bf 63}, 124015 (2001)
  [arXiv:gr-qc/0101052].


\bibitem{Bir}
D.~Birmingham,
  Phys.\ Rev.\  D {\bf 64}, 064024 (2001)
  [arXiv:hep-th/0101194].

\bibitem{LKM}
  H.~W.~Lee, Y.~W.~Kim and Y.~S.~Myung,
  arXiv:0807.1371 [hep-th].


\bibitem{KN}
  Y.~Kwon and S.~Nam,
  arXiv:1001.5106 [hep-th].



\bibitem{Set}
  M.~R.~Setare,
  Class.\ Quant.\ Grav.\  {\bf 21}, 1453 (2004)
  [arXiv:hep-th/0311221].

\bibitem{LM}
  Y.~S.~Myung and H.~W.~Lee,
  Mod.\ Phys.\ Lett.\  A {\bf 21}, 1737 (2006)
  [arXiv:hep-th/0506031].


\end{thebibliography}
\end{document}